\documentclass[a4paper, 10pt]{article}
\usepackage[dvips]{graphicx}

\usepackage{amsthm}
\usepackage{amsmath}

\author{P.H. Lundow\footnote{KTH Physics, AlbaNova University Center,
SE-106 91 Stockholm, Sweden, \texttt{phl@kth.se}.} and
K. Markstr\"om\footnote{Department of Mathematics and Mathematical
Statistics, Ume\aa{} University, SE-901 87 Ume\aa, Sweden,
\texttt{klas.markstrom@math.umu.se}.}}

\title{Reconstruction of the finite size canonical ensemble from
  incomplete micro-canonical data}

\theoremstyle{plain}

\theoremstyle{definition}

\theoremstyle{remark}

\newcommand{\dd}{\,\mathrm{d}}
\newcommand{\FZ}{\mathcal{Z}}
\newcommand{\FF}{\mathcal{F}}
\newcommand{\FU}{\mathcal{U}}
\newcommand{\FC}{\mathcal{C}}
\newcommand{\FS}{\mathcal{S}}
\newcommand{\GS}{\Gamma_1}
\newcommand{\GK}{\Gamma_2}
\newcommand{\prob}[1]{\mathsf{Pr}\left(#1\right)}
\newcommand{\avg}[1]{\langle #1 \rangle}
\newcommand{\var}[1]{\mathsf{Var}\left(#1\right)}
\newcommand{\ulo}{u}
\newcommand{\uhi}{v}
\newcommand{\pow}[1]{\!\cdot\!10^{#1}}
\newcommand{\amu}{\bar{\mu}}
\newcommand{\achi}{\bar{\chi}}

\begin{document}
  \maketitle

  \begin{abstract}
      In this paper we discuss how partial knowledge of the density of
      states for a model can be used to give good approximations of
      the energy distributions in a given temperature range.  From
      these distributions one can then obtain the statistical moments
      corresponding to eg the internal energy and the specific heat.
      These questions have gained interest apropos of several recent
      methods for estimating the density of states of spin models.
      
      As a worked example we finally apply these methods to the
      3-state Potts model for cubic lattices of linear order up to
      128. We give estimates of eg latent heat and critical
      temperature, as well as the microcanonical properties of
      interest.
  \end{abstract}


\section{Introduction}\label{sec:introduction}

When studying a statistical mechanical model the most complete
information is given by the density of states function.  From complete
knowledge of the density of states one can immediately work with the
microcanonical ensemble and of course also compute the partition
function and through it have access to the canonical ensemble as well.
The main problem here is that computing the density of states for
systems of even very modest size is typically very hard.  However,
 recently several sampling schemes which strive to approximate the
density of states have appeared.  One recent method was given
\cite{wang-landau:01} and in \cite{wangswen} several such methods were
given, and in \cite{sampart} all of the later methods as well as several
others were united in a common framework.

For work in the microcanonical ensemble the mentioned methods give all
the information needed. Using them one can find the density of states
in an energy interval around the critical region and that is all that 
is needed for most investigations of the critical properties of the
model. The microcanonical ensemble is more refined than the canonical 
ensemble in that every equilibrium measure for the canonical ensemble 
is found among the equilibrium measures for the microcanonical
ensemble, but for some models there are microcanonical equilibria
which are not present in the canonical ensemble. For a fuller survey of
the mathematical theory of ensemble equivalence see \cite{ellis:04} and
its references. This means that all properties of the thermodynamic
limit can be obtained via the microcanonical ensemble.

However, even in view of what has been said the canonical ensemble has
its own interest for finite systems.  Among other things it governs the
behaviour of many sampling algorithms and for systems where we have
ensemble nonequivalence its dynamic can be very interesting.  In order
to reconstruct the canonical ensemble one would in principle need to
know the density of states for all values of the energy $E$.  However,
 using methods as in \cite{sampart} this is very costly, and also not
needed for work in the microcanonical ensemble.

Our aim is to look at how density of states data from a restricted
interval of energies can be used to get an approximation of the energy
distribution of the canonical ensemble for some range of couplings
$K$.  Thanks to the strong concentration of the energy distributions
we will see that one can obtain a very good approximation of the
energy distribution and through its moments most of the standard
thermodynamical properties.  This will be demonstrated first in a case
where we know the exact partition function, the Ising model on the
$256\times 256$ square lattice, and then for a case where we have
ensemble nonequivalence: the 3-state potts model on the 3-dimensional
cubic lattice.  All in all we find that with data collected with the
methods of \cite{sampart} in mind one can get a good picture of the
canonical ensemble as well as the microcanonical.  In fact, thanks to
knowing the density of states for a full interval of energies we will
be able to reconstruct the canonical ensemble for all couplings in
some interval rather than just those used in the sampling process.

\section{Notation}\label{sec:notation}
  Let us define what we need in terms of the Ising model. Later on,
  when the Potts model is our subject, we will redefine some
  quantities, but our general discussion will be held in terms of the
  Ising model. Let $G$ be a graph on $n$ vertices $V=\{1,\ldots,n\}$
  and $m$ edges. A state is a function $s : V \to Q$ where
  $Q=\{+1,-1\}$ and we say that vertex $i$ has spin $s_i$. The energy
  of a state is defined as $E(s) = \sum_{ij}s_is_j$ where the sum is
  taken over all edges $ij$ of the graph and we have $-m \le E \le
  m$. The magnetisation of $s$ is defined as $M(s) = \sum_i s_i$ so
  that $-n\le M\le n$.

  A normalised energy and magnetisation will often be used, here
  defined as $U=E/m$ and $\mu=M/n$ so that $-1\le U, \mu \le 1$.  The
  number of states having energy $E$ and magnetisation $M$ is denoted
  $a(E,M)$. The number of states at energy $E$, or, the density of
  states, is denoted $a(E)$, where, of course, $a(E)=\sum_M
  a(E,M)$. 

  From quotients of $a(E)$ we obtain what we will refer to as the
  coupling function
  \[K(U) = \frac{1}{\ell}\,\log\frac{a(E)}{a(E+\ell)}\]
  where $U=E/m$ and $\ell$ is the difference between two consecutive
  energies.  The very fundamental entropy function
  \[S(U) = \frac{\log a(E)}{n}\]
  is of course related to the coupling function through
  \[K(U) = -\frac{n}{m}S'(U)\]
  See \cite{sampart} for proofs and further details.

  The partition function is defined for all graphs as \[\FZ(K,H) =
  \sum_{E,M} a(E,M)\,\exp(K\,E+H\,M)\] where $K$ and $H$ are the
  dimensionless coupling and external field respectively.  When the
  external field is zero we simplify as 
  \[\FZ(K) = \sum_E a(E)\,\exp(K\,E)\]  
  As a convention we will write our coupling dependent quantities 
  in a calligraphic font, such as $\FZ(K)$.
  
  The central moments of a random variable $X$ are defined
  \[
  \sigma_i = \avg{\left(X-\avg{X}\right)^i}, \quad i=0,1,\ldots
  \]
  where $\sigma_0=1$, $\sigma_1=0$ and $\sigma_2=\var{X}$.  The
  standard deviation is written $\sigma=\sqrt{\sigma_2}$. The cumulants 
  $\kappa_i$ of a distribution, or, the $i$:th derivatives of $\log\FZ$, 
  can be expressed in terms of moments.  For the first few we have
  \begin{align*}
    \frac{\partial   \log \FZ(K)}{\partial K  } = & \kappa_1 = \avg{E} \\
    \frac{\partial^2 \log \FZ(K)}{\partial K^2} = & \kappa_2 = \var{E} = 
    \sigma_2 \\
    \frac{\partial^3 \log \FZ(K)}{\partial K^3} = & \kappa_3 = \sigma_3 \\
    \frac{\partial^4 \log \FZ(K)}{\partial K^4} = & \kappa_4 = 
    \sigma_4 - 3\,\sigma_2^2
  \end{align*}
  The free energy is here defined as
  \[
  \FF(K) = \frac{1}{n}\,\log\FZ(K)
  \]
  and the reader should note that we have used a simplified version
  compared to its traditional form. The internal energy, specific heat 
  and coupling dependent entropy are given by 
  \[
  \FU(K) = \frac{1}{m}\, \frac{\partial \log\FZ(K)}{\partial K} =
  \frac{1}{m}\, \langle E \rangle.
  \]
  \[
  \FC(K) = \frac{1}{m}\, \frac{\partial^2 \log\FZ(K)}{\partial K^2}
  = \frac{1}{m}\, \var{E}.
  \]
  \[
  \FS(K) = \FF(K) - \frac{m}{n}\,K\,\FU(K)
  \]
  We would also like to study the higher derivatives in the form of
  skewness
  \[
  \GS(K) = \frac{
    \partial^3 \log\FZ(K) / \partial K^3
  }{
    \left(\partial^2 \log\FZ(K) / \partial K^2\right)^{3/2}
  } = \frac{\sigma_3}{\left(\sigma_2\right)^{3/2}}
  = \frac{\kappa_3}{\left(\kappa_2\right)^{3/2}}
  \]
  and (excess) kurtosis
  \[
  \GK(K) = \frac{
    \partial^4 \log\FZ(K) / \partial K^4
  }{
    \left(\partial^2 \log\FZ(K) / \partial K^2\right)^2
  } = \frac{\sigma_4}{\sigma_2^2} - 3
  = \frac{\kappa_4}{\kappa_2^2}
  \]
  Note that for normal distributions they both evaluate to zero.

  Derivatives with respect to the field $H$ are of course obtained
  analogously. The magnetisation and susceptibility are defined
  respectively as
  \begin{gather*}
    \mu(K,H) = \frac{1}{n}\, \frac{\partial \log\FZ(K,H)}{\partial H} =
    \frac{1}{n}\,\avg{M} \\
    \chi(K,H) = \frac{1}{n}\, \frac{\partial^2 \log\FZ(K,H)}{\partial H^2}
    = \frac{1}{n}\, \var{M}
  \end{gather*}
  However, what one usually want is the spontaneous magnetisation and
  susceptibility. As finite size approximations of these we use
  \begin{gather*}
    \amu(K) = \frac{1}{n}\,\avg{|M|} \\ 
    \achi(K) = \frac{1}{n}\,\var{|M|} 
  \end{gather*}
  and assume that these converge to the appropriate limits.  

  Given a lattice of side $L$ with $L^3$ vertices we call $L$ the
  linear order of he lattice.  When necessary we will subscript the
  functions with the linear, as in $\FZ_L$.

\section{Distributions of energy}

  In this section we will look at how the distribution of energies for
  a given coupling $K_0$ can be reconstructed.  The process is rather
  straightforward and follows more or less by definition, but we will
  derive it in some detail. We will first derive an exact expression for 
  $\mathrm{Pr}(E)$ when $S'(U)$ is exactly known, next we look at
  what can be done when only partial knowledge of $K(U)$ is available,
  and finally we consider precision issues for such incomplete
  reconstructions.
 

\subsection{From coupling to distribution}
 Our first aim is to express $\mathrm{Pr}(E)$ in terms of the values
 of $K(U)$ in some interval of energies $\ulo\le U \le \uhi$.

  We assume the Boltzmann distribution for the states, that is, if we
  sample at a coupling $K_0$ the probability for our system being in
  state $s$ is
  \[\prob s = \frac{\exp(K_0\,E(s))}{\FZ(K_0)}\]
  and consequently the probability for our system being in a state of
  energy $E$ is
  \begin{equation}\label{eq:pre1}
    \prob E = \frac{a(E)\,\exp(K_0\,E)}{\FZ(K_0)}
  \end{equation}
  Recall that we defined $a(E) = \exp(n\,S(U))$. Then we obtain
  \begin{equation}\label{eq:pre2}
    \prob E = \frac{\exp\left( n\,S(U) + m\,U\,K_0\right)}{\FZ(K_0)}
  \end{equation}
  By definition we also have
  \[
  S(U) = \int_{-1}^{U} \!\! S'(x) \dd x = \int_{-1}^{\ulo} \!\! S'(x)
  \dd x + \int_{\ulo}^{U} \!\! S'(x) \dd x = A_{u} -
  \frac{m}{n}\,\int_{\ulo}^{U} \!\!  K(x) \dd x
  \]
  and trivially
  \[
  U = \ulo + \int_{\ulo}^U 1 \dd x
  \]
  Plugging these identities into Equation~\ref{eq:pre2} and applying
  only a modicum of algebraic manipulation it simplifies finally into
  \begin{equation}\label{eq:pre3}
    \prob E = C_{u}\,\exp\left( m\,\int_{\ulo}^{U}\!\! K_0 - K(x) \dd x \right)
  \end{equation}
  Since the outcome is a probability function the constant $C_{u}$ can
  also be defined by normalising so that
  \begin{equation}\label{eq:pre4}
    \sum_E \prob E = 1
  \end{equation}
  We will consider $C_{u}$ further in the next section.  Finally, we
  note in passing that the derivative of the probability function with
  respect to $U$ is $m\,(K_0 - K(U))\,\prob E$.  Thus the points where
  the sign of the derivative changes is determined by when $K_0 =
  K(U)$.

  Note that we have only defined the function $K(U)$ at discrete
  points $U=E/m$ so we should be somewhat careful with how the
  integral is taken.  If a function $f(x)$ is defined at $a=x_0<x_1<
  \ldots < x_p=b$ then we use a left-point rule for integration
  \[
  \int_a^bf(x)\dd x = \sum_{i=0}^{p-1} f(x_i)\,(x_{i+1}-x_i)
  \]

  Having reconstructed the distribution of energies the moments
  and cumulants are easily retrieved. First the average
  \[
    \avg{E} = \sum_E E\,\prob E
  \]
  and then the central moments
  \[
  \sigma_i = \sum_E \left(E - \avg{E}\right)^i\,\prob E
  \]
  and from these we obtain the sought-after estimates of the
  derivatives by evaluating the cumulants of the distribution so that,
  for example
  \[
  \kappa_4 = \sigma_4 - 3\,\sigma_2^2 = \frac{\partial^4\log\FZ}{\partial K^4}
  \]
  
  Let us address the issue of derivatives with respect to the field as
  well. If we during our sampling process remembered to collect data
  on the magnetisation as well, then we can reconstruct the
  spontaneous magnetisation and susceptibility as well. Our program
  should then collect raw moments on the form $\avg{|M|^i\,\mid
  E}$. Then the following holds
  \begin{gather}
    \amu(K_0) = \frac{1}{n}\,\avg{|M|} =
    \frac{1}{n}\,\avg{\avg{|M|\,\mid E}} = \frac{1}{n}\, \sum_E
    \avg{|M|\,\mid E}\,\prob E \label{eq:mag}\\
    \achi(K_0) = \frac{1}{n}\,\var{|M|} = \frac{1}{n}\, \left(\sum_E 
    \avg{|M|^2\,\mid E}\,\prob E -
    \left(\sum_E \avg{|M|\,\mid E}\,\prob
    E\right)^2\right)\label{eq:sus}
  \end{gather}
  The following is a nice alternative way of writing the variance
  \[
  \var{|M|} = \avg{\var{|M|\,\mid E}} + \var{\avg{|M|\,\mid E}}
  \]
  that is, the variance is the sum of the expectation of the variances
  and the variance of the expectations.

\subsection{Reconstruction from incomplete data.}  
In the previous subsection we assumed that $K(U)$ was exactly known
for all energies.  When this is the case we have seen that
$\mathrm{Pr}(E)$ can be exactly determined, as one would expect.  Let
us now assume that our data contains information on quotients of
consecutive density of states, or rather, that we have available
estimates for the coupling function $K(U)$ for an interval of energies
$\ulo\le U \le \uhi$.

In this situation we can no longer determine $\mathrm{Pr}(E)$
completely from Equation~\ref{eq:pre3} since we can no longer compute $C_{u}$, 
and we may have errors in our estimate for $K(E)$. However, if the
expected energy lies well within the interval $\ulo\le U \le \uhi$ we 
can hope that a good enough approximation can be found, due to the
rapid decay of the tails of  the energy distribution.

The simplest way to find such an approximation is simply to let
Equation \ref{eq:pre4} define an approximate value $C_{u}'$ via
\begin{equation}\label{aeq}
      \sum_{ \ulo\le E/m \le \uhi } \mathrm{Pr}_{u}( E )= 1,
\end{equation}
where $\mathrm{Pr}_{u}( E )$ is obtained by using $C_{u}'$ instead of $C_{u}$ in
Equation~\ref{eq:pre3}. Equation \ref{aeq} is simply a linear equation
for $C_{u}'$ and so can be easily solved. Once $C_{u}'$ has been found 
we can use $C_{u}'$ and our estimated $K(U)$ in Equation \ref{eq:pre3}
to compute an approximate distribution function for the canonical
ensemble.

\subsection{Accuracy of the reconstructed distributions}
  Since the canonical ensemble is always determined by the density of 
  states we only have two sources of errors: the precision of the
  original data and the truncation error due to not having data from all
  energies.

  In a perfect world the collected data comes from the entire interval
  of energies $-1\le U\le 1$. However, normally it suffices for the
  interval to be wide enough to cover the energies at coupling $K_0$
  with a high probability.  In short, the distribution of energies
  corresponding to $K_0$ must stay in the interval $[\ulo,\uhi]$ with
  a probability close to $1$.  If $[\ulo,\uhi]$ only covers say, 99\%
  or less of the energies you see at $K_0$, the normalising step in 
  Equation~\ref{eq:pre4} will produce erroneous results.

  Given a coupling $K_0$ that is close to the critical coupling $K_c$
  we expect the distribution to be anything but normal (ie gaussian).
  But, as we move away from $K_c$ the distribution typically becomes
  close to normal.  For example, at $K=0$ the distribution is clearly
  approaching a normal one with increasing system size.  It has been
  shown, see \cite{martinlof:73}, that this also holds for Ising
  systems when $K$ is greater than some $K_1>K_c$.  Since our approach
  is somewhat pragmatic we will only assume that far enough from $K_c$
  the energy distributions can be treated as roughly normal.

  For how large or small values of $K_0$ does the procedure return a
  credible distribution on $[\ulo,\uhi]$? Treating the distribution as
  roughly normal it should be enough, for all practical purposes, to
  make sure that the end-points are at least $4$, if possible $5$ and
  preferably $6$ standard deviations $\sigma$ away.

\begin{figure}[!hbt]
    \begin{center}
      \includegraphics[width=0.99\textwidth]{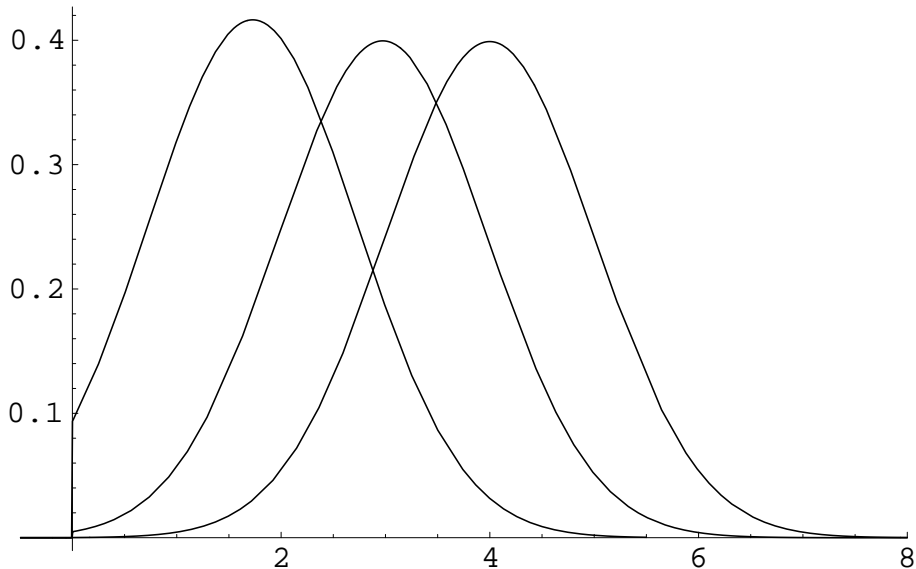}
    \end{center}
    \caption{Cut-off normal distributions such that
    $\avg{X}=k\,\sigma$ for $k=2,3,4$.}
    \label{fig:cutnormal}
  \end{figure}

  The probability density of a normally distributed variable is
  \[f(x)=\frac{1}{\sqrt{2\pi}}\,\exp(-x^2/2)\]  
  We will translate it slightly to the right, ie take $f(x-p)$ for
  $p>0$, and cut it off at $x=0$. Now let $X$ be a random variable
  with probability density
  \[
  g(x) = \left\{\begin{array}{ll}
  \frac{f(x-p)}{A(p)} & x>0 \\
  0                & x\le 0
  \end{array}\right.
  \]
  where $A(p)$ is the mass of probability on $x>0$, ie
  \[
  A(p) = \int_0^{\infty} f(x-p) \dd x
  \]
  so that $g(x)$ becomes a cut-off, but otherwise normal looking,
  probability density on the real axis. For which $p$ is the y-axis,
  ie the cut-off point, located $k$ standard deviations $\sigma$ away?
  Numerical calculations gave Table~\ref{tab:cutnormal1} below and in
  Figure~\ref{fig:cutnormal} the distribution functions are shown for
  $k=2,3,4$. In the Table we also list the errors
  \[
  \epsilon_i = \left| \kappa_i(g) - \kappa_i(f)\right|
  \]
  that is, we take the difference in cumulants for the cut-off density
  $g$ and the normal density $f$ translated $k$ standard deviations.
  These errors are of course very idealised, being based on normal
  distributions, and should be considered rough guidelines. For any
  particular distribution we will see different errors and especially
  the higher cumulants will deviate from these. 
  \begin{table}[!hbt]
    \begin{center}
      \begin{tabular}{|lccccc|}
	\hline
	$k$&$p$ &$\epsilon_1$ &$\epsilon_2$ &$\epsilon_3$ &$\epsilon_4$   \\
	\hline
	2 &1.728042230 &$2\pow{-1}$ &$2\pow{-1}$ &$2\pow{-1}$ &$2\pow{-1}$\\
	3 &2.973669402 &$2\pow{-2}$ &$1\pow{-2}$ &$4\pow{-2}$ &$8\pow{-2}$\\
	4 &3.998789790 &$1\pow{-3}$ &$5\pow{-4}$ &$2\pow{-3}$ &$7\pow{-3}$\\
	5 &4.999979927 &$2\pow{-5}$ &$7\pow{-6}$ &$4\pow{-5}$ &$2\pow{-4}$\\
	6 &5.999999885 &$6\pow{-7}$ &$4\pow{-8}$ &$2\pow{-7}$ &$1\pow{-6}$\\
	\hline
      \end{tabular}
      \caption{Peak location $p$ and cumulant errors for a cut-off
      normal distribution with $\avg{X}=k\,\sigma$.}
      \label{tab:cutnormal1}
    \end{center}
  \end{table}

  \section{The 2-dimensional Ising model}
  We will employ the 2D Ising model as a test bed for our method.
  Recall that the critical coupling is $K_c =
  \mathrm{arctanh}\,(\sqrt{2}-1) \approx 0.4407$ and that the critical
  energy is $U_c = 1/\sqrt{2} \approx 0.7071$.  In \cite{exactart} we
  computed the exact partition function for the $256 \times
  256$-lattice with periodic boundary. However, since the largest
  density of states $a(0)$ has $19726$ digits we will take the liberty
  of doing all actual computations with 50 digits numerical precision
  instead.

\begin{figure}[!hbt]
    \begin{minipage}{0.49\textwidth}
      \begin{center}
	\includegraphics[width=0.99\textwidth]{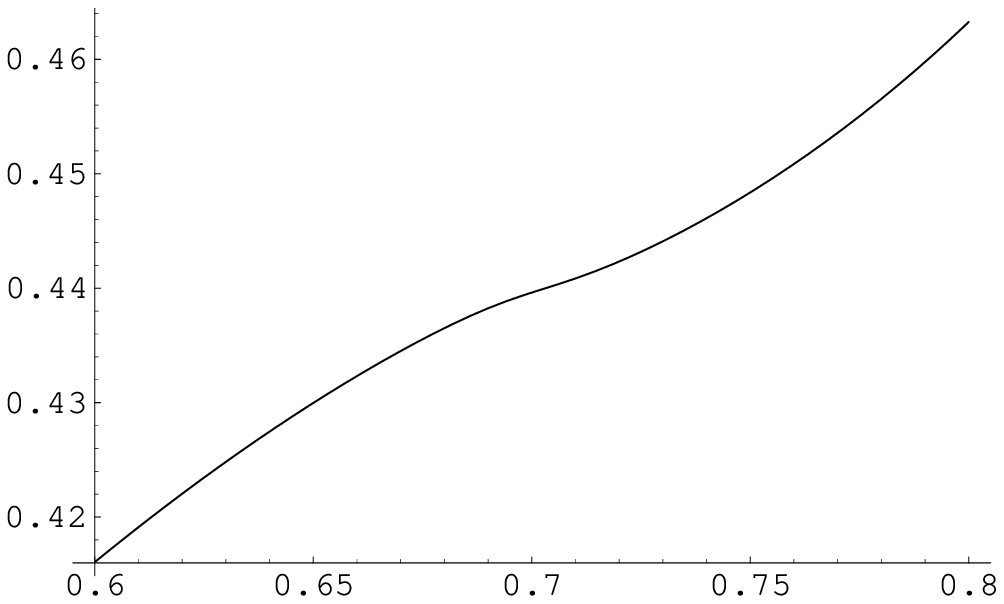}
      \end{center}
    \end{minipage}%
    \begin{minipage}{0.49\textwidth}
      \begin{center}
	\includegraphics[width=0.99\textwidth]{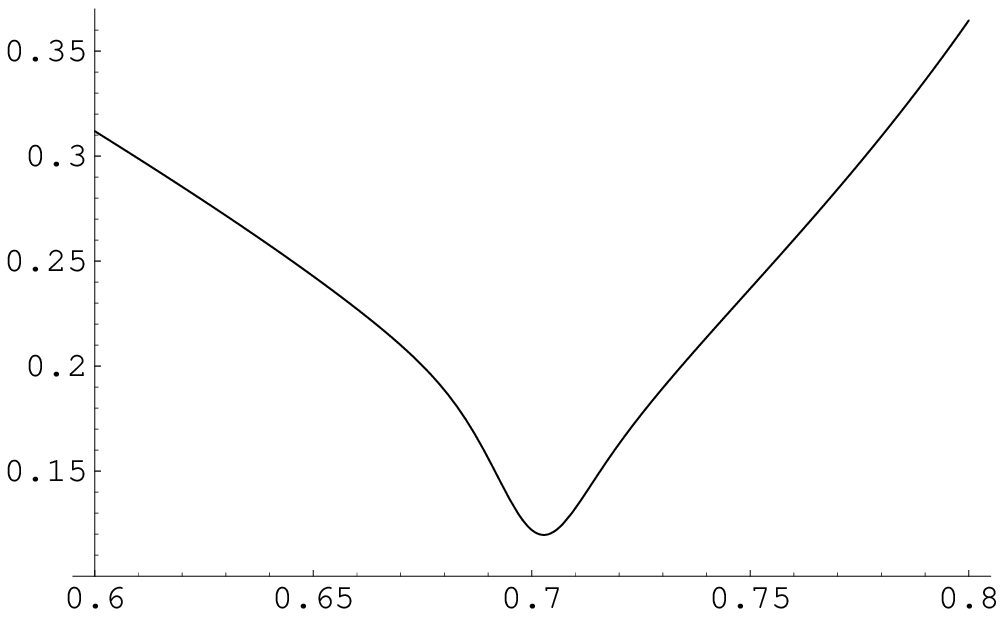}
      \end{center}
    \end{minipage}
    \caption{$K(U)$ and $K'(U)$ for $256\times 256$-lattice.}
    \label{fig:K-ising}
  \end{figure}

\begin{figure}[!hbt]
    \begin{center}
      \includegraphics[width=0.99\textwidth]{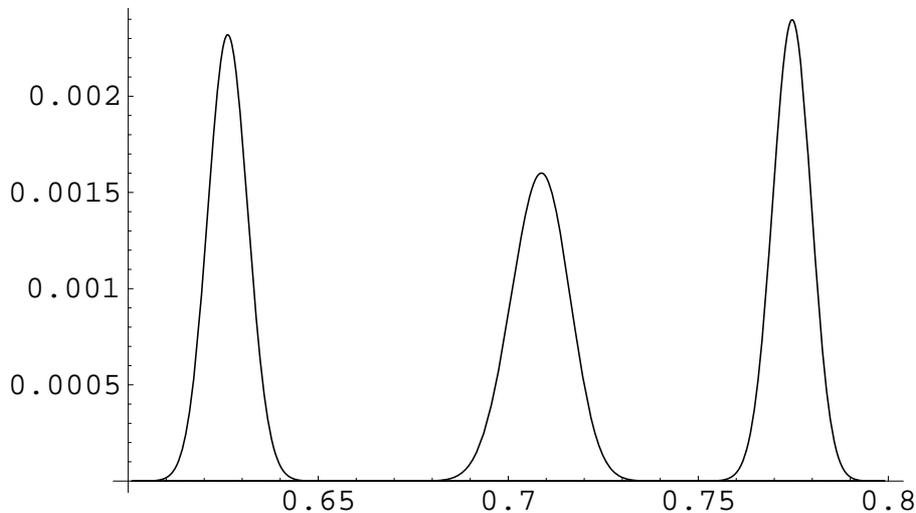}
    \end{center}
    \caption{Probability densities of energies at $K_1=0.423780$
    (left), $K_c$ (middle) and $K_2=0.454942$ (right) for $256\times
    256$-lattice. At $K_1$ and $K_2$ the interval bounds are $5\sigma$
    away. Probability $\prob E$ on the y-axis and energy $U=E/m$ on
    the x-axis.}
    \label{fig:dens-ising}
  \end{figure}

  Suppose now that we have collected data on $K(U)$ for $\ulo=0.6 \le
  U \le 0.8=\uhi$, an interval comprising $6554$ energies. In
  Figure~\ref{fig:K-ising} we plot $K(U)$ and $K'(U)$ for $L=256$.
  From the exact (50 digits) coupling function on the interval
  $[0.6,0.8]$ we reconstruct the distribution, ie the probability
  density function, of energies at $K_c$ using
  Equation~\ref{eq:pre3}. Let $\epsilon_i$ denote the relative error
  of the $i$:th cumulant where we compare the cumulant $\kappa_i$ of
  the reconstructed energy distribution with the $i$:th derivative of
  $\log\FZ$, ie
  \[
  \epsilon_i = \left|\frac{\kappa_i}{\partial^i \log\FZ/\partial K^i}
  - 1\right|
  \]
  The relative errors $\epsilon_1, \epsilon_2, \epsilon_3, \epsilon_4$
  are negligibly small, less than $1\pow{-30}$.  However, this
  distribution lives clearly in the middle of our energy interval
  $[0.6,0.8]$, the lower bound being $14\sigma$ below and the upper
  bound $12\sigma$ above the mean.  In Table~\ref{tab:2d-example1} we
  compute relative errors of the cumulants when the coupling
  corresponds to a cut-off distribution with $\avg{E}$ located
  $k\,\sigma$ from the lower bound $\ulo=0.6$ for $k=2,3,4,5,6$ and in
  Table~\ref{tab:2d-example2} we do the corresponding at the other end
  of the interval so that $\avg{E}$ is $k\,\sigma$ from the upper bound
  $\uhi=0.8$.  Figure~\ref{fig:dens-ising} shows the probability
  densities at $K=0.423780$, $K=K_c$ and $K=0.454942$.

  \begin{table}[!hbt]
    \begin{center}
      \begin{tabular}{|lccccc|}
	\hline 
	$k$ & $K$ &$\epsilon_1$ &$\epsilon_2$ &$\epsilon_3$ &$\epsilon_4$\\ 
	\hline
	2 & 0.418736 &$7\pow{-4}$ &$2\pow{-1}$ &$1\pow{+1}$&$9\pow{+1}$\\
	3 & 0.420644 &$4\pow{-5}$ &$1\pow{-2}$ &$1\pow{0}$ &$4\pow{+1}$\\
	4 & 0.422226 &$9\pow{-7}$ &$4\pow{-4}$ &$6\pow{-2}$&$2\pow{0}$ \\
	5 & 0.423780 &$7\pow{-9}$ &$4\pow{-6}$ &$8\pow{-4}$&$4\pow{-2}$\\
	6 & 0.425338 &$2\pow{-11}$&$1\pow{-8}$ &$3\pow{-6}$&$2\pow{-4}$\\
	\hline
      \end{tabular}
      \caption{Relative errors of cumulants for cut-off distribution
      with $\avg{X}=\ulo+k\,\sigma$.}
      \label{tab:2d-example1}
    \end{center}
  \end{table}

  \begin{table}[!hbt]
    \begin{center}
      \begin{tabular}{|lccccc|}
	\hline 
	$k$ & $K$ &$\epsilon_1$ &$\epsilon_2$ &$\epsilon_3$ &$\epsilon_4$\\
	\hline
	2 & 0.460385 &$5\pow{-4}$ &$2\pow{-1}$ &$6\pow{0}$  &$5\pow{+1}$\\
	3 & 0.458331 &$2\pow{-5}$ &$1\pow{-2}$ &$8\pow{-1}$ &$2\pow{+1}$\\
	4 & 0.456623 &$5\pow{-7}$ &$3\pow{-4}$ &$3\pow{-2}$ &$1\pow{0}$ \\
	5 & 0.454942 &$4\pow{-9}$ &$3\pow{-6}$ &$3\pow{-4}$ &$1\pow{-2}$\\
	6 & 0.453254 &$7\pow{-12}$&$7\pow{-9}$ &$9\pow{-7}$ &$4\pow{-5}$\\
	\hline
      \end{tabular}
      \caption{Relative errors of cumulants for cut-off distribution
      with $\avg{X}=\uhi-k\,\sigma$.}
      \label{tab:2d-example2}
    \end{center}
  \end{table}

  We also computed the cumulant errors at $K_c$ with the upper and
  lower bound of the energy interval located $6\,\sigma$ away.  For
  $L=32, 64, 128, 256$ the errors are quite small, $\epsilon_1 <
  1\pow{-12}$, $\epsilon_2 < 2\pow{-10}$, $\epsilon_3 < 2\pow{-8}$ and
  $\epsilon_4< 6\pow{-8}$.  For $L\leq 16$, the errors become larger,
  but on the other hand for such small graphs it is easy to collect a
  complete set of $K(U)$-data instead of only a short interval.

\begin{figure}[!hbt]
    \begin{minipage}{0.49\textwidth}
      \begin{center}
	\includegraphics[width=0.99\textwidth]{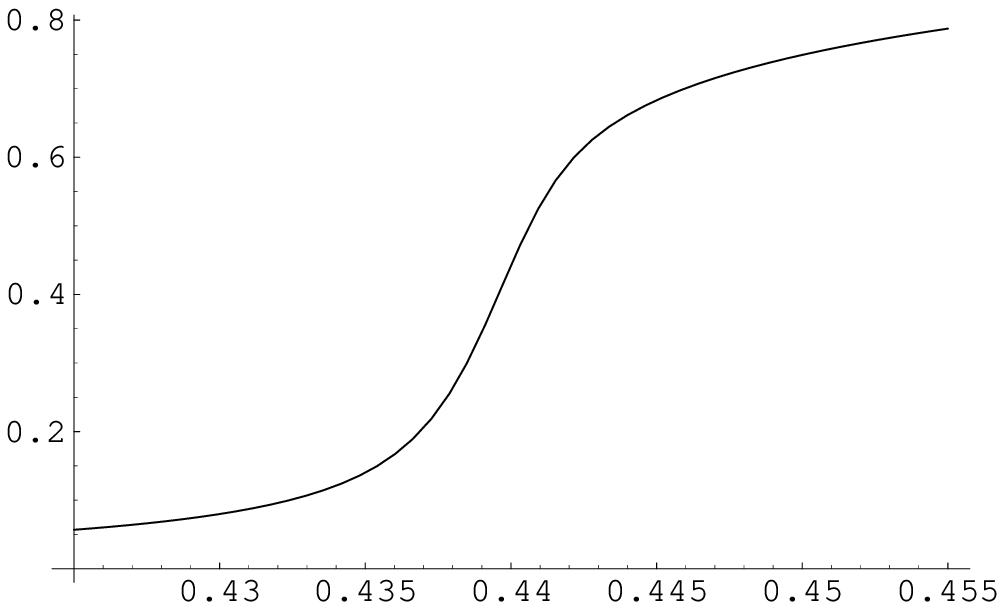}
      \end{center}
    \end{minipage}%
    \begin{minipage}{0.49\textwidth}
      \begin{center}
	\includegraphics[width=0.99\textwidth]{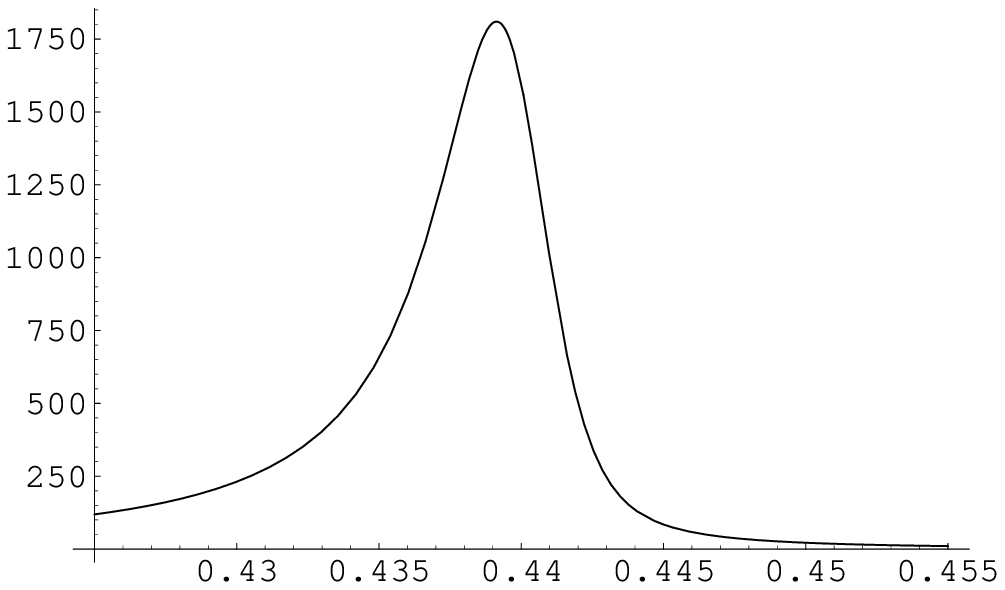}
      \end{center}
    \end{minipage}
    \caption{$\amu(K)$ and $\achi(K)$ for $256\times 256$-lattice.}
    \label{fig:MX-ising}
  \end{figure}

  Regarding the magnetisation and susceptibility we have no way of
  comparing the reconstructed values with exact values.  We have
  simply run the Metropolis method at 10 different temperatures in the
  vicinity of $K_c$ ($0.42\le K\le 0.46$) and collected magnetisation
  moments at each energy level.  Using these data and the exact
  $K$-function we can reconstruct the magnetisation and susceptibility
  at any temperature in that region using Equation~\ref{eq:mag} and
  \ref{eq:sus} adding data as prescribed in \cite{sampart}.  See
  Figure~\ref{fig:MX-ising} for a snap-shot of the result.  The
  reconstructed curves agrees very well with the sampled data.

  \section{The free energy}
  By definition we have that
  \[
  \FF(K) = \FF(0) + \frac{m}{n}\,\int_0^K \FU(x) \dd x
  \]
  where the constant $\FF(0) = \log 2$ for the Ising model, and
  $\FF(0) = \log q$ for the $q$-state Potts model. Having evaluated
  $\FU(K)$ for a number of values of $K$ this is of course easily
  accomplished.  Unfortunately, this formulation implies that we have
  collected data so that the energy distribution can be reconstructed
  for $K\ge 0$.  For smaller graphs this is of course perfectly
  alright but for large graphs this was exactly what we wanted to
  avoid. However, due to the well-behaved nature of the internal
  energy $\FU(K)$ we can circumvent this problem. Suppose we have
  reconstructed the internal energy for two system sizes $L_1$ and
  $L_2$, where $L_1 < L_2$. Suppose further that we have
  $\FU_{L_1}(K)$ for $0 \le K \le b_1$ and $\FU_{L_2}(K)$ for $a \le K
  \le b_2$ where $0 \le a \le K_c \le b_2 \le b_1$.  Then, for $a\le K
  \le b_2$ we have
  \[
  \FF(K) = \FF(0) + \frac{m}{n}\int_0^a \FU_{L_1}(x) \dd x + \frac{m}{n}
  \int_a^K \FU_{L_2}(x) \dd x + \epsilon
  \]
  where $\epsilon$ is an error term.  

  How big is the error? Let $f$ and $g$ be continuous functions on the
  interval $[a,c]$ with $a<b<c$. Then the following elementary
  calculation gives an estimate:
  \begin{gather*}
    \int_a^c g(x) \dd x = \int_a^b g(x) \dd x + \int_b^c g(x) \dd x = \\
    = \int_a^b g(x)-f(x)+f(x) \dd x + \int_b^c g(x) \dd x = \\
    = \int_a^b f(x) \dd x + \int_b^c g(x) \dd x + \epsilon
  \end{gather*}
  where $\epsilon$ is the error term
  \[\epsilon = \int_a^b g(x) - f(x) \dd x \]
  which gives the very simple but useful estimate
  \begin{equation}\label{eq:error1}
    |\epsilon| \le (b-a)\,\max_{a\le x\le b}|g(x)-f(x)|
  \end{equation}
  Since the internal energy function is an increasing and, in fact,
  convex function, it is easy to establish the maximum. The
  integration is numerical so it is important to evaluate $\FU(K)$ at
  points chosen densely enough, with special attention to values close
  to $K_c$ where $\FU(K)$ is expected to change rapidly.

  \subsection{A worked example for the 2D Ising model}
  Here our goal is to compute the free energy at $K=K_c$ for the
  $256\times 256$ 2D Ising model by using a sequence of system sizes,
  $L=32, 64, 128, 256$, and formulate the method as
  \begin{gather*}
    \FF_{256}(K_c) = \log 2 + 2 \int_0^{0.15}\!\FU_{32}(x) \dd x +
    2\int_{0.15}^{0.30}\!\FU_{64}(x) \dd x \,+ \\
    + 2\int_{0.30}^{0.40}\!\FU_{128}(x) \dd x +
    2\int_{0.40}^{K_c}\!\FU_{256}(x) \dd x
  \end{gather*}
  though the integration will of course be numerical.  To evaluate the
  $\FU_L(x)$ at the couplings indicated by the integral boundaries we
  use the exact (to 50 digits precision) $K$-functions at intervals
  wide enough to keep the endpoints $6\,\sigma$ away for each energy
  distribution. Picking simple values gives the following energy
  intervals, coupling intervals and step lengths used for evaluating
  $\FU_L(K)$.
  \[
  \begin{array}{lrll}
    L=32,  &-0.15\le U \le 0.30, & 0.00 \le K \le 0.15, & h=0.0010 \\
    L=64,  &0.05\le U \le 0.45,  & 0.15 \le K \le 0.30, & h=0.0005 \\
    L=128, &0.30\le U \le 0.60,  & 0.30 \le K \le 0.40, & h=0.0002 \\
    L=256, &0.50\le U \le 0.75,  & 0.40 \le K \le 0.441,& h=0.0001
  \end{array}
  \]
  The intervals for $K$ were chosen to give good overlaps, and are of
  course model dependent.
  
  Numerical integration of these data points, using eg the trapezoidal
  method gives that $\FF_{256}(K_c) \approx 0.92970521$. The correct
  answer is $\FF_{256}(K_c) = 0.92970516\ldots$ giving an error of 
  $5.5\pow{-8}$.

  The error contribution given by Equation~\ref{eq:error1} is only of
  the order of $6\pow{-12}$. The main error source is actually the
  numerical integration. As is well-known, numerical evaluation of
  $\int_a^b f(x)\dd x$ with the trapezoidal rule gives the error term
  \[
  \epsilon = -\frac{(b-a)\,h^2}{12}\,f''(\xi), \quad a<\xi<b
  \]
  where $h$ is the step length.  Since the function we integrate is
  $\FU(K)$ its second derivative is
  \[
  \FU''(K) = \frac{1}{m}\,\frac{\partial^3 \log \FZ(K)}{\partial K^3}
  = \frac{\kappa_3}{m}
  \]
  and it is at a very little extra cost we evaluate the third cumulant
  when we already have the distribution.
  
  For example, the error contributed from
  \[2\int_{a=0}^{b=0.15} \FU_{32}(x) \dd x\]
  is at most 
  \[
  \frac{2\,(b-a)\,h^2}{12} \max_{a<x<b} \FU''_{32}(x) =
  \frac{2\,(0.15-0)\,0.001^2}{12} 1.67 \approx 4.2\pow{-8}
  \]
  and the errors contributed by the other integrals are at most
  respectively $3.4\pow{-8}$ for $L=64$, $1.6\pow{-8}$ for $L=128$ and
  $5.8\pow{-8}$ for $L=256$ and they sum up to $1.5\pow{-7}$ which is
  clearly larger than the actual error we received.

  \section{An example with a first order phase transition: The
  3-dimensional 3-state Potts model}
  
  For this model we need to redefine some of our quantities.  A state
  is here a function $s : V \to Q$ where $Q$ is a set of $q$ distinct
  elements, eg $Q=\{1,\ldots,q\}$. The energy is defined as $E(s) =
  \sum_{ij} \delta(s_i,s_j)$, where $\delta(x,y)$ is the
  Kronecker-delta, so that $0\le E \le m$.  We normalise as before and
  let $U=E/m$ so that $0\le U \le 1$. Let $\eta_j=\sum_i
  \delta(s_i,j)$, ie the number of vertices having spin $j$. For the
  Potts model, the definition of magnetisation $M$ varies slightly in
  the literature. For example, $M=\max(\eta_1,\ldots,\eta_q)$ is
  sometimes used, but $M=\eta_1^2 + \ldots + \eta_q^2$ is the one used
  here. This means that $n^2/q \le M \le n^2$ and we normalise by
  taking $\amu=\mu=M/n^2$ so that $1/q \le \amu \le 1$.  Having
  defined these quantities their physical versions follow accordingly.

  \subsection{The sampled data}
  
  The data were generated and collected by using the sampling method
  described in detail in \cite{sampart} and we refer to that paper for
  further details.  Since this model is conjectured to have a first
  order phase transition and cluster methods thus are expected to have
  exponential mixing time \cite{slow} we opted for a highly optimised
  single spin Metropolis method.  We used up to a few hundred independent
  spin systems, which after slowly being brought to the right coupling
  were given a few days or weeks, depending on their size, of
  continuous running for mixing.  The length of the sampling runs were
  of the same order. The sampled also passed the consistency test from
  \cite{sampart}.
  
  For the smaller lattices we collected at least 10000 measurements
  per energy level, often orders of magnitude more.  For the larger
  lattices, say $L\ge 32$, this quickly becomes difficult.  For
  $L=96,128$ we did not manage to fill out all energy levels inside
  the energy jump at critical the coupling, though these empty levels
  are \emph{very} few relatively speaking.  From these data we then
  constructed the coupling function $K(U)$.
  
\begin{figure}[!hbt]
    \begin{minipage}{0.49\textwidth}
      \begin{center}
	\includegraphics[width=0.99\textwidth]{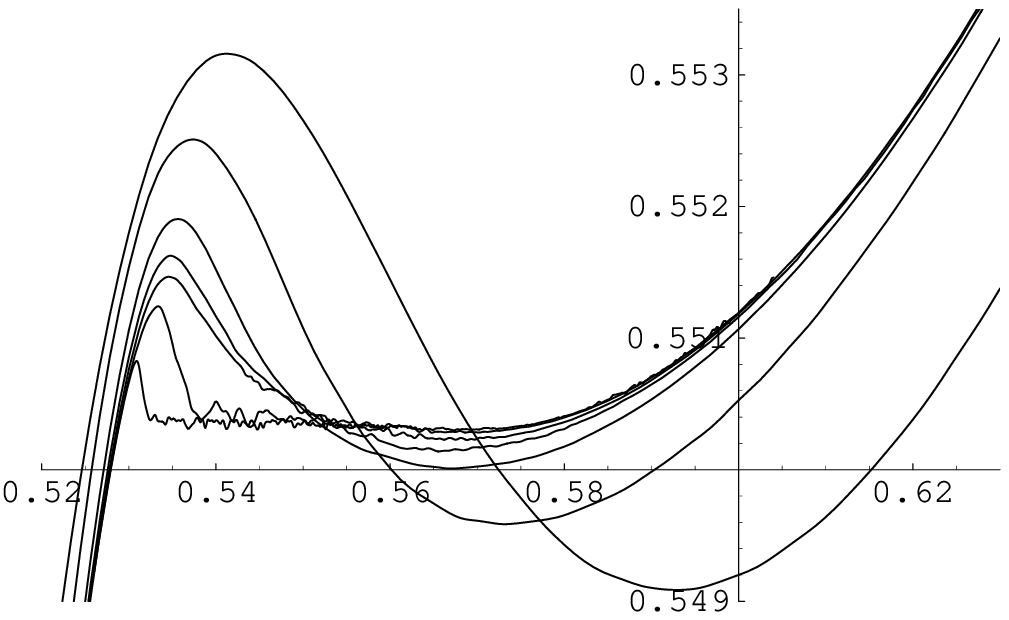}
      \end{center}
    \end{minipage}%
    \begin{minipage}{0.49\textwidth}
      \begin{center}
	\includegraphics[width=0.99\textwidth]{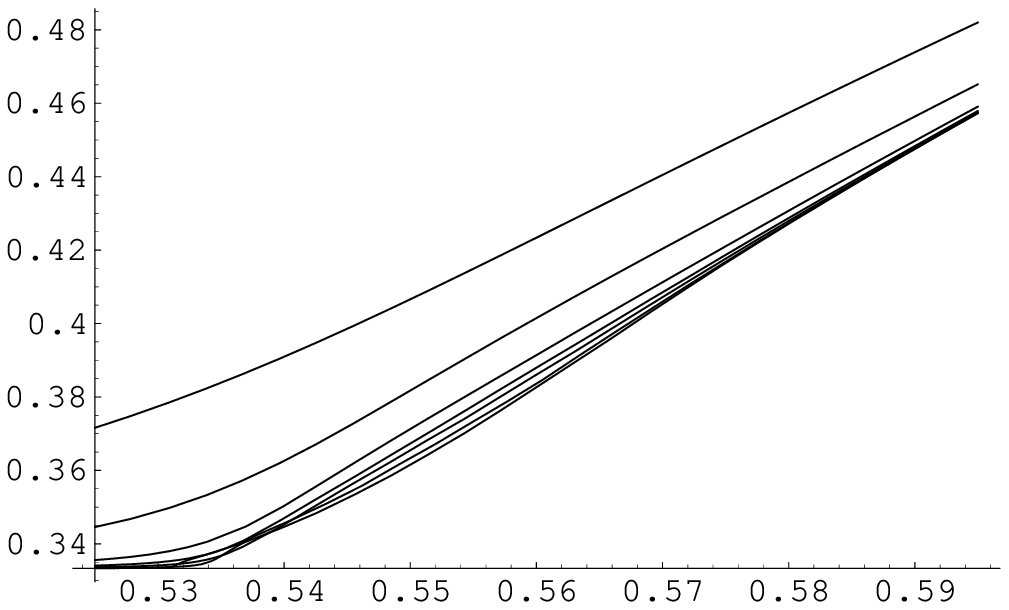}
      \end{center}
    \end{minipage}
    \caption{$K(U)$ (left) and $\mu(U)$ (right) for $L\ge 16$.}
    \label{fig:6}
  \end{figure}

  The coupling functions $K(U)$, from which the energy distributions
  are generated, are shown in the left plot of Figure~\ref{fig:6} and
  the right plot shows the magnetisations $\mu(U)$.  Note that the
  $K$-functions behave rather different from that of the Ising model
  in Figure~\ref{fig:K-ising}.  Here the $K$-functions have their own
  set of critical points and they are listed in Table~\ref{tab:2}.
  Let $U^-$ and $U^+$ be the locations of the maximum and the minimum
  respectively of $K(U)$. The corresponding values of $K$ at these
  points are denoted $K^-$ and $K^+$ respectively. We define the
  latent heat as $U^{\pm} = U^+ - U^-$. Let also $\mu^+$ and $\mu^-$
  denote the magnetisation at respectively $U^+$ and $U^-$.
  \begin{table}[!hbt]
    \begin{center}
      \begin{tabular}{|r|ccccccc|}
	\hline
	$L$ &$U^-$&$U^+$&$U^{\pm}$&$K^-$&$K^+$&$\mu^-$&$\mu^+$ \\
	\hline
	6&   0.54502& 0.60677& 0.06174& 0.553398& 0.548924& 0.4174& 0.5125 \\
	8&   0.54138& 0.59245& 0.05101& 0.553159& 0.549088& 0.3930& 0.4779 \\
	12&  0.53851& 0.57979& 0.04128& 0.552756& 0.549367& 0.3694& 0.4442 \\
	16&  0.53721& 0.57327& 0.03606& 0.552507& 0.549588& 0.3580& 0.4264 \\
	24&  0.53603& 0.56884& 0.03281& 0.552133& 0.549861& 0.3474& 0.4119 \\
	32&  0.53564& 0.56717& 0.03153& 0.551905& 0.550012& 0.3427& 0.4057 \\
	48&  0.53478& 0.56645& 0.03167& 0.551625& 0.550151& 0.3380& 0.4012 \\
	64&  0.53457& 0.56796& 0.03340& 0.551465& 0.550233& 0.3363& 0.4029 \\
	96&  0.53337& 0.56816& 0.03479& 0.551242& 0.550284& 0.3344& 0.4019 \\
	128& 0.53090& 0.56785& 0.03694& 0.550826& 0.550306& 0.3336& 0.4004 \\
	\hline
      \end{tabular}
    \end{center}
    \caption{Critical points and values of the combinatorial
    functions.}
    \label{tab:2}
  \end{table}
  
\subsection{The reconstructed ensemble}
  
  Next we used the $U$-dependent functions from the previous section to
  reconstruct the $K$-dependent quantities in the way described earlier in
  the paper.  Let us here stress that none of the functions plotted here
  were sampled directly, ie we did not keep track of the variance and
  expectation of the energy in the sampling runs and everything here
  is based on the microcanoncial data.
  
\begin{figure}[!hbt]
    \begin{minipage}{0.49\textwidth}
      \begin{center}
	\includegraphics[width=0.99\textwidth]{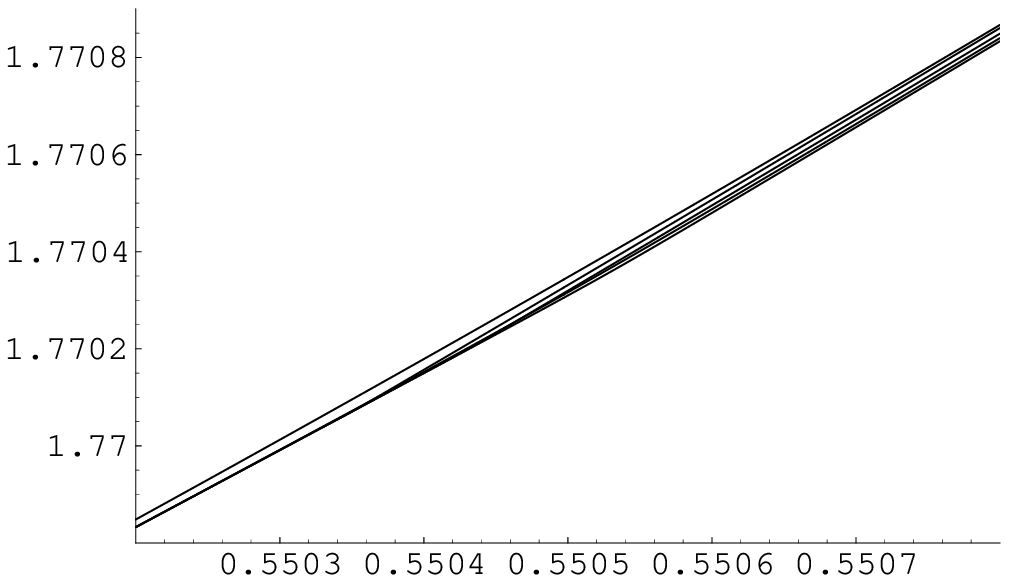}
      \end{center}
    \end{minipage}%
    \begin{minipage}{0.49\textwidth}
      \begin{center}
	\includegraphics[width=0.99\textwidth]{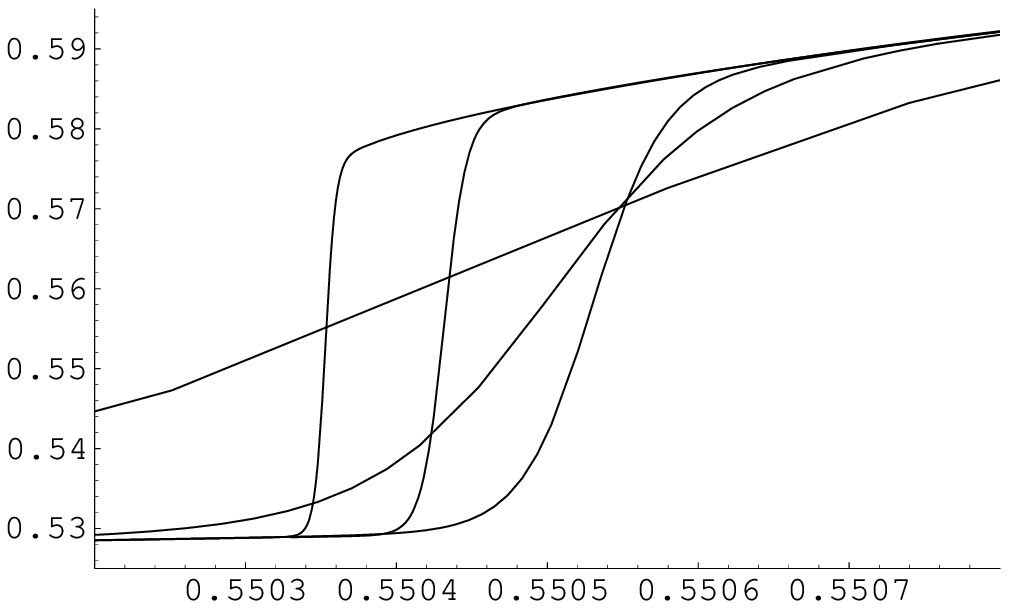}
      \end{center}
    \end{minipage}
    \caption{Free energy $\FF(K)$ (left) and internal energy $\FU(K)$
    (right) for $L\ge 32$.}
    \label{fig:1}
  \end{figure}

\begin{figure}[!hbt]
    \begin{minipage}{0.49\textwidth}
      \begin{center}
	\includegraphics[width=0.99\textwidth]{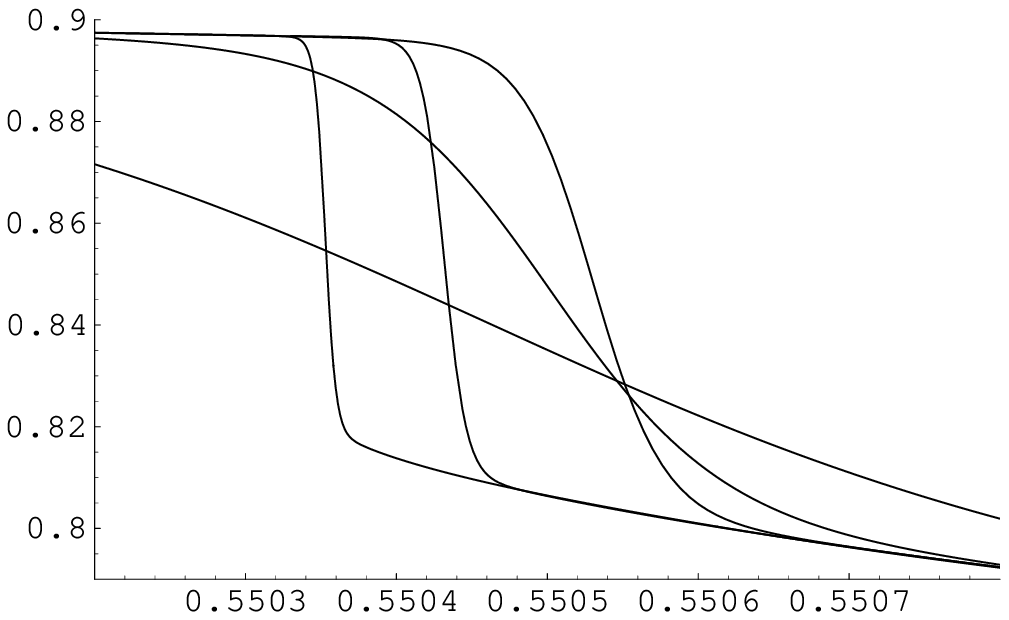}
      \end{center}
    \end{minipage}%
    \begin{minipage}{0.49\textwidth}
      \begin{center}
	\includegraphics[width=0.99\textwidth]{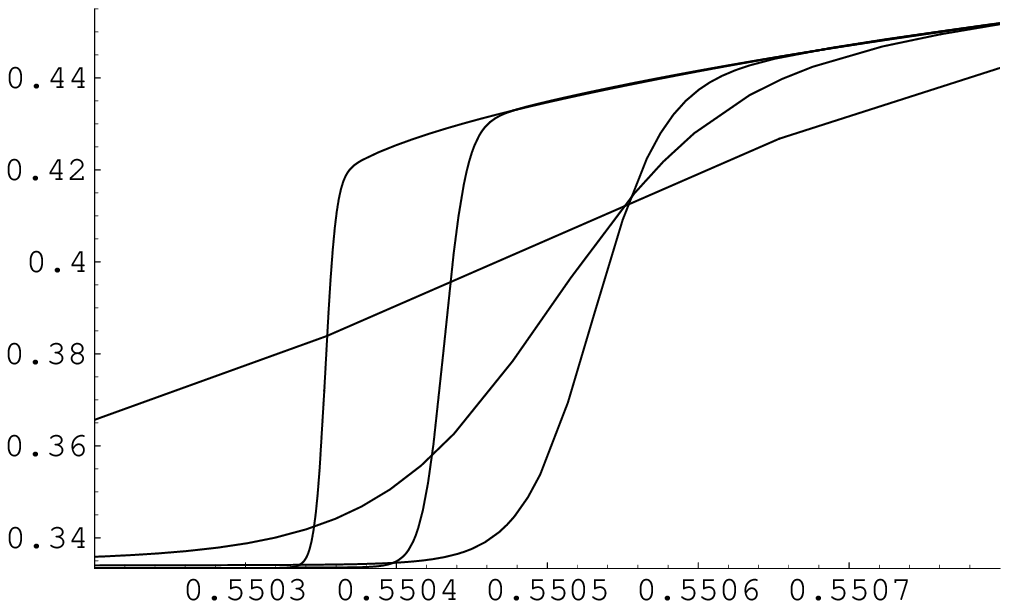}
      \end{center}
    \end{minipage}
    \caption{Entropy $\FS(K)$ (left) and magnetisation $\mu(K)$
    (right) for $L\ge 32$.}
    \label{fig:2}
  \end{figure}

  First we will show a quick gallery of pictures of the physical, ie
  coupling dependent, quantities that were defined in
  Section~\ref{sec:notation}.  In Figure~\ref{fig:1} we show the free
  energy $\FF$ and the internal energy $\FU$ in a narrow region around
  $K_c$ for the larger lattices.  The dramatic jump in the energy of
  course leads to a similar behaviour in the entropy $\FS(K)$ shown in
  Figure~\ref{fig:2}. This is also seen in the magnetisation $\mu(K)$
  in the same figure. We find that everything agrees well with the
  expected first order nature of the phase transition.


\begin{figure}[!hbt]
    \begin{minipage}{0.49\textwidth}
      \begin{center}
	\includegraphics[width=0.99\textwidth]{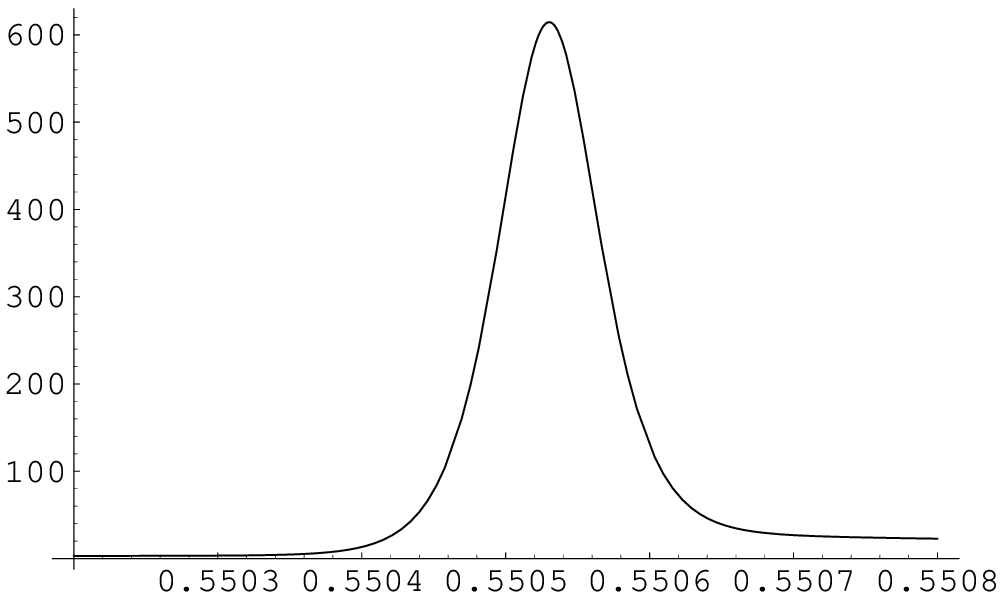}
      \end{center}
    \end{minipage}%
    \begin{minipage}{0.49\textwidth}
      \begin{center}
	\includegraphics[width=0.99\textwidth]{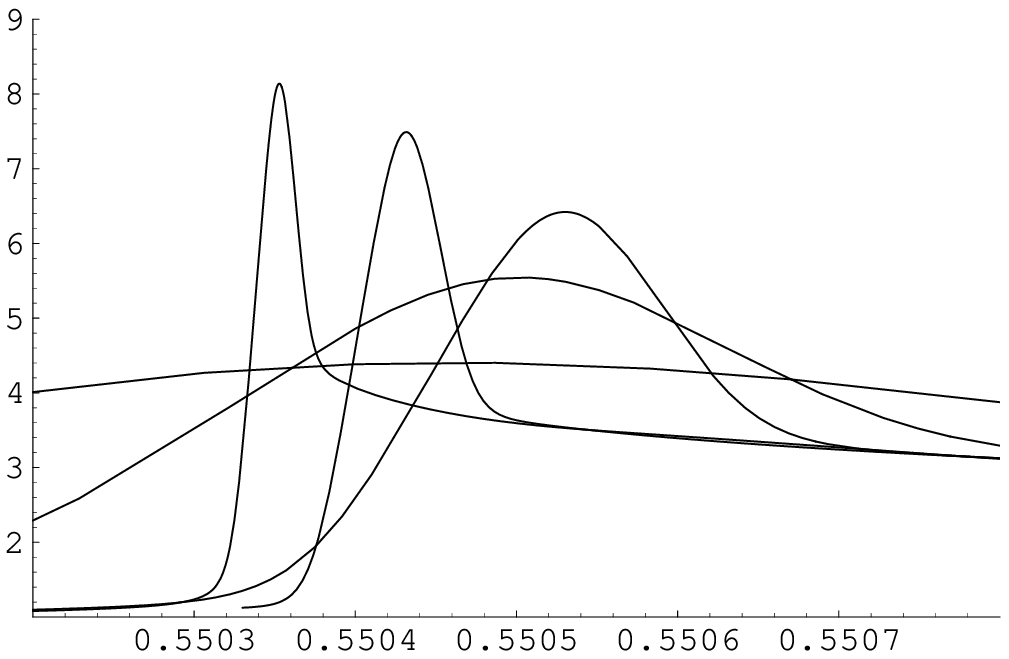}
      \end{center}
    \end{minipage}
    \caption{Specific heat $\FC(K)$ for $L=64$ (left) and its
    logarithm for $L\ge 32$ (right).}
    \label{fig:3}
  \end{figure}

\begin{figure}[!hbt]
    \begin{minipage}{0.49\textwidth}
      \begin{center}
	\includegraphics[width=0.99\textwidth]{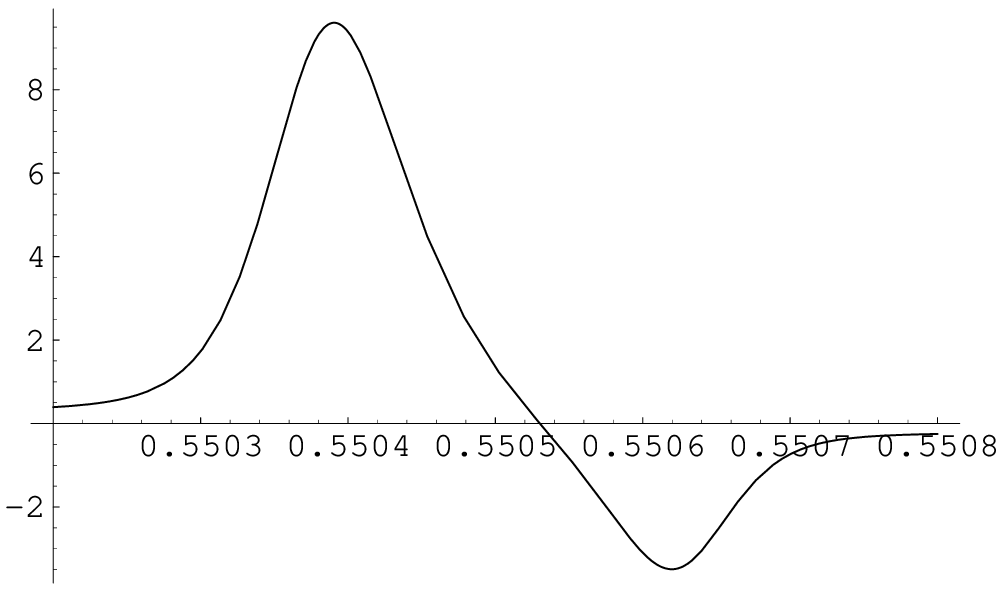}
      \end{center}
    \end{minipage}%
    \begin{minipage}{0.49\textwidth}
      \begin{center}
	\includegraphics[width=0.99\textwidth]{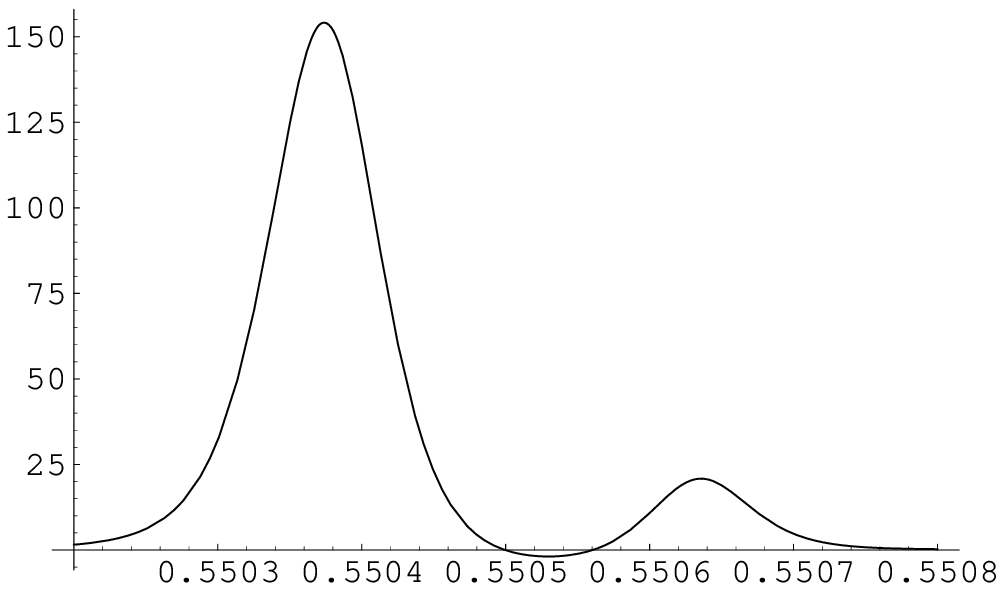}
      \end{center}
    \end{minipage}
    \caption{Skewness $\Gamma_1(K)$ (left) and kurtosis $\Gamma_2(K)$
      (right) for $L=64$.}
    \label{fig:4}
  \end{figure}

\begin{figure}[!hbt]
    \begin{minipage}{0.49\textwidth}
      \begin{center}
	\includegraphics[width=0.99\textwidth]{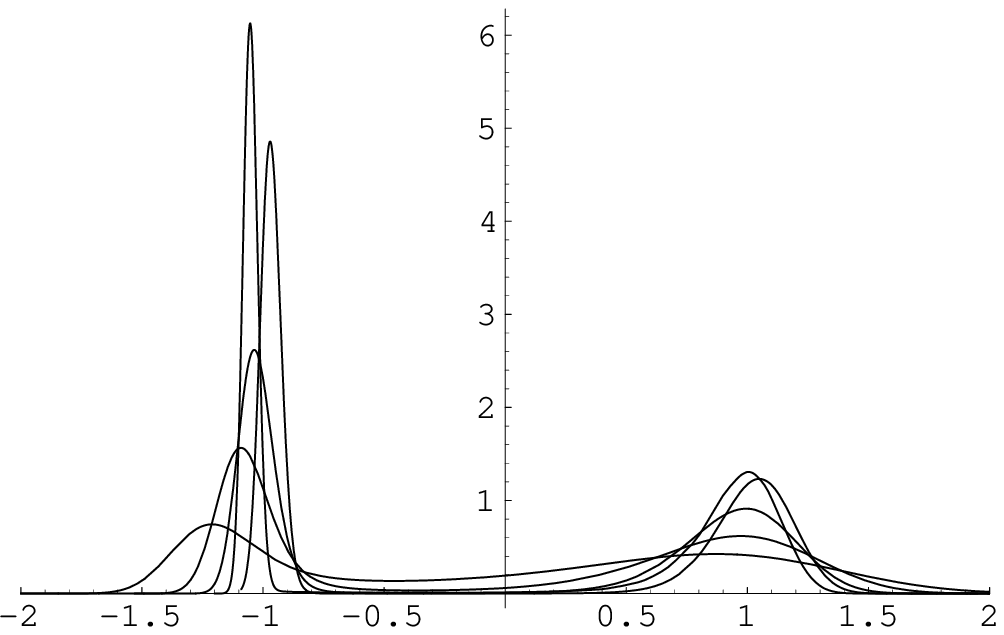}
      \end{center}
    \end{minipage}%
    \begin{minipage}{0.49\textwidth}
      \begin{center}
	\includegraphics[width=0.99\textwidth]{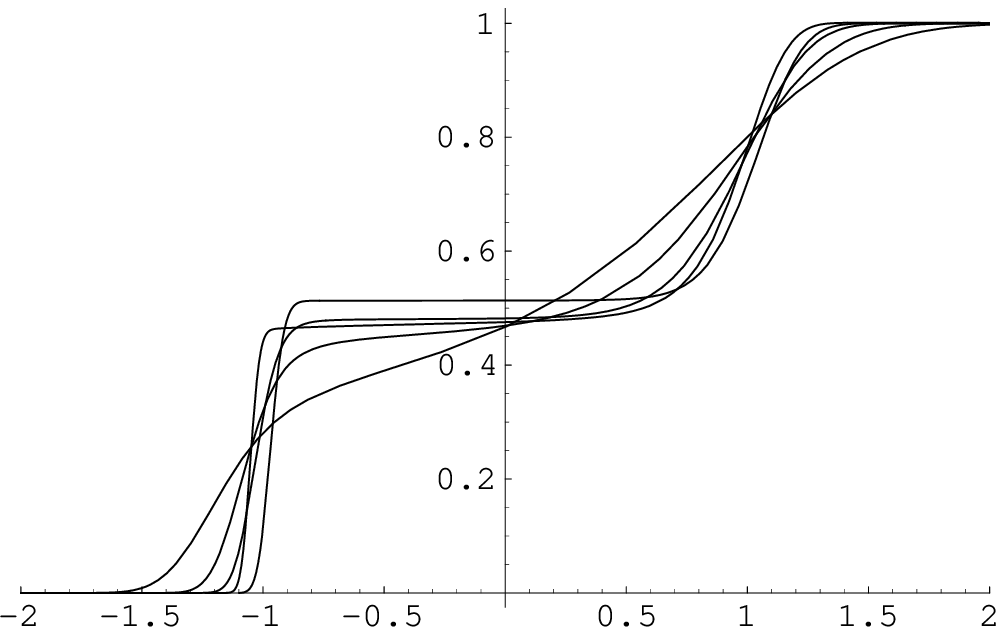}
      \end{center}
    \end{minipage}
    \caption{Normalised probability densities (left) and
      distribution functions (right) vs $\sigma$ at $K^*$
      for $L\ge 32$.}
    \label{fig:5}
  \end{figure}

  The specific heat $\FC(K)$ is shown for $L=64$ in the left plot of
  Figure~\ref{fig:3}. The maximum of this quantity grows very fast
  with $L$ and to be able to compare them for several $L$ the right
  plot shows their logarithm.
  This can also be said about the skewness and kurtosis in
  Figure~\ref{fig:4}. These quantities changes sign in a number of
  critical points so taking logarithms is not advisable.
  The distributions go through a sharply bimodal phase as the coupling
  moves past $K_c$. Define $K^*$ as the point where the specific heat
  has its maximum. What do the distributions at this point look like?
  In the left plot of Figure~\ref{fig:5} the probability densities
  $p(x)$ of the normalised variable $x=(E-\avg{E})/\sigma(E)$ at $K^*$
  are shown, while the right plot shows the distribution functions
  (accumulated densities) defined as
  \[
  \Phi(x) = \int_{-\infty}^x p(t) \dd t.
  \]
  Note, by the way, that the peaks in the probability densities are
  very close $\pm \sigma$.
  In Table~\ref{tab:1} the data connected with $K^*$ are listed.
  \begin{table}[!hbt]
    \begin{center}
      \begin{tabular}{|r|cccccc|}
	\hline
	$L$ &$K^*$&$\FC(K^*)$&$\FU(K^*)$&$\FF(K^*)$&$\FS(K^*)$&$\mu^*$ \\
	\hline
	6  & 0.555045 & 5.1279 & 0.63766 & 1.78444 & 0.72265 & 0.5557 \\
	8  & 0.552821 & 7.2889 & 0.61575 & 1.77697 & 0.75577 & 0.5130 \\
	12 & 0.551143 & 12.539 & 0.59190 & 1.77220 & 0.79355 & 0.4643 \\
	16 & 0.550665 & 19.584 & 0.57983 & 1.77093 & 0.81306 & 0.4381 \\
	24 & 0.550460 & 41.929 & 0.56831 & 1.77034 & 0.83185 & 0.4118 \\
	32 & 0.550462 & 81.880 & 0.56318 & 1.77028 & 0.84026 & 0.3996 \\
	48 & 0.550502 & 255.97 & 0.55921 & 1.77032 & 0.84678 & 0.3901 \\
	64 & 0.550530 & 614.58 & 0.55821 & 1.77036 & 0.84842 & 0.3877 \\
	96 & 0.550432 & 1790.0 & 0.55550 & 1.77020 & 0.85291 & 0.3821 \\
	128& 0.550353 & 3422.7 & 0.55279 & 1.77008 & 0.85738 & 0.3763 \\
	\hline
      \end{tabular}
    \end{center}
    \caption{Critical points $K^*$ and values of the physical
    quantities.}
    \label{tab:1}
  \end{table}
\subsection{Asymptotics}
\begin{figure}[!hbt]
      \begin{minipage}{0.49\textwidth}
	\includegraphics[width=0.99\textwidth]{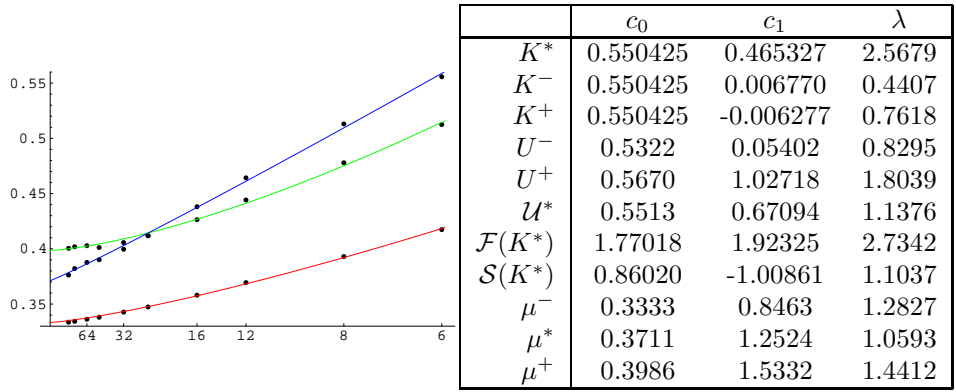}
      \end{minipage}%
      \begin{minipage}{0.49\textwidth}
	      \begin{center}
		\begin{tabular}{|r|ccc|}
		  \hline
		  
		   & $c_{0}$&$c_{1}$&$\lambda$ \\
		  \hline
		  $K^*$      & 0.550425& 0.465327 & 2.5679 \\
		  $K^-$      & 0.550425& 0.006770 & 0.4407 \\
		  $K^+$      & 0.550425&-0.006277 & 0.7618 \\
		  $U^-$      & 0.5322  & 0.05402  & 0.8295 \\
		  $U^+$      & 0.5670  & 1.02718  & 1.8039 \\
		  $\FU^* $   & 0.5513  & 0.67094  & 1.1376 \\
		  $\FF(K^*)$ & 1.77018 & 1.92325  & 2.7342 \\
		  $\FS(K^*)$ & 0.86020 & -1.00861 & 1.1037 \\
		  $\mu^-$    & 0.3333  & 0.8463   & 1.2827 \\
		  $\mu^*$    & 0.3711  & 1.2524   & 1.0593 \\
		  $\mu^+$    & 0.3986  & 1.5332   & 1.4412 \\
		  \hline
		\end{tabular}
	      \end{center}
	      \label{tab:3}

      \end{minipage}%
    \caption{Left: magnetisations $\mu^+$, $\mu^*$, $\mu^-$ (downwards
      at y-axis) vs $1/L$. Right: coefficients for our fitted curves} 
  \end{figure}

  In this section we will see how some of the values in the tables
  above scale with the linear order. First we wish to establish the
  critical coupling $K_c$. We have three separate sequences of
  critical points which should all converge to $K_c$, namely $K^*$,
  $K^+$ and $K^-$. The coefficients of the fits described below 
  are collected in Table \ref{tab:3}.

  The sequences $K^+$ and $K^-$ from Table~\ref{tab:2} have the nice
  feature that they are monotone; $K^+$ is increasing and $K^-$ is
  decreasing. Unfortunately though, the $K^-$-sequence appears
  slightly blemished for $L=128$, as is the $U^-$-sequence.  Even so,
  after discarding that particular point and assuming that the
  sequences stay monotone also for larger $L$ we then have upper and
  lower bounds on $K_c$. Then $0.550306 \le K_c \le 0.551242$, a
  rather wide interval. The story is different for the $K^*$-sequence,
  it sometimes increases, sometimes decreases, but we see nothing
  amiss with the value for $L=128$.

  To establish a $K_c$ we have attempted a simple fit of the form $c_0
  + c_1\,L^{-\lambda}$ for some coefficients $c_0, c_1$ and exponent
  $\lambda$ to the data for $K^*$.  To find the parameters we used
  Mathematica's non-linear fitting function.  Our fitting function
  applied to this sequence gives acceptable fits.  We will try to
  estimate the error in this approach by fitting a curve to data of
  the form $L_{\min}\le L \le 128$ with $L_{\min}=6,8,12$, ie for
  three sets of data.  Leaving out more points gives the fitted curve
  an unconvincing look.  This gave
  \[
  K_c = 0.550425 \pm 0.000025
  \]
  which agrees with the previous interval. This estimate is a little
  lower than that of \cite{janke-villanova:97} (who also provide a
  nice table of previous results) but their data is based on rather
  small graphs, $L\le 36$. On the other hand, our estimate ends up
  right in the middle of the (rather wide) interval given by
  \cite{guttman-enting:94}.

\begin{figure}[!hbt]
    \begin{minipage}{0.49\textwidth}
      \begin{center}
	\includegraphics[width=0.99\textwidth]{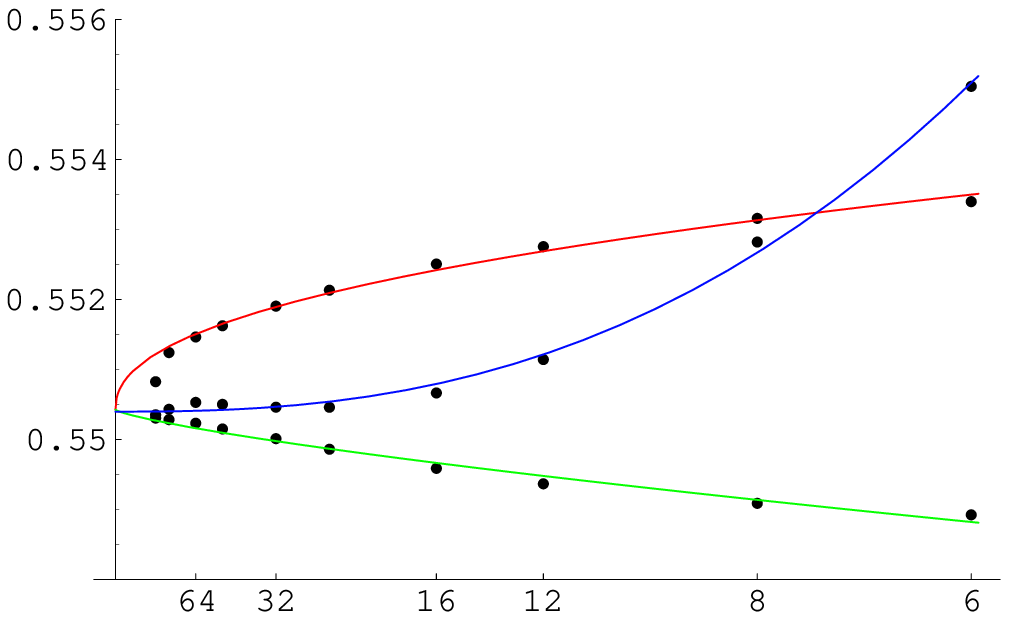}
      \end{center}
    \end{minipage}%
    \begin{minipage}{0.49\textwidth}
      \begin{center}
	\includegraphics[width=0.99\textwidth]{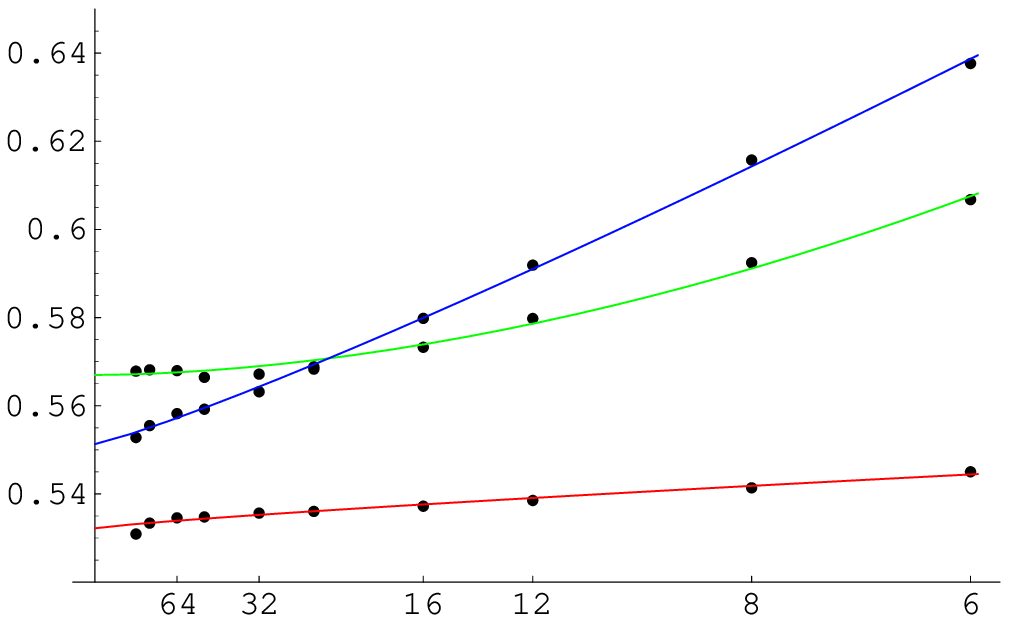}
      \end{center}
    \end{minipage}
    \caption{Left: couplings $K^-$, $K^*$, $K^+$ (downwards) vs
      $1/L$. Right: energies $U^+$, $U^*$, $U^-$ (downwards at y-axis)
      vs $1/L$.}
    \label{fig:scale-1}
  \end{figure}

  From the interval above we choose the mid-point as our limit, ie we
  set $K_c=c_0=0.550425$, and fit all points to determine the
  remaining parameters.  Using the same limit we fitted curves to the
  $K^-$ (discarding $L=128$) and $K^+$ data.  We received the curves
  shown with the points in the left plot of Figure~\ref{fig:scale-1}.

  A different behaviour is expected from the three sequences of
  energies; $U^+$, $U^-$ and $U^*$. Here $U^+\to U^+_c$ and $U^-\to
  U^-_c$ and the difference $U^{\pm}=U^+-U^-$ should converge to the
  latent heat $U^{\pm}_c$,  whereas $\FU^*$ should converge to some 
  value $\FU_c$ between $U^-$ and $U^+$.  Again we see a possibly
  too big jump in the data for $U^-$ at $L=128$ so we will discard
  this point. Applying the process we described above we received
  \begin{gather*}
    U^-_c = 0.5322 \pm 0.0013 \\
    U^+_c = 0.5670 \pm 0.0004 \\
    \FU_c = 0.5513 \pm 0.0008 
  \end{gather*}
  Again, using the mid-points as limits we received the curves shown
  with the data points in the right plot of Figure~\ref{fig:scale-1}.
  Using these estimates of $U^+_c$ and $U^-_c$ also provides us with
  estimates of the asymptotic latent heat $U^{\pm}$.  This resulted in
  \[
  U^{\pm}_c = U^+_c - U^-_c = 0.03480 \pm 0.0017
  \]
  which is clearly smaller than the estimate $0.0538$ (after division
  by 3) of \cite{janke-villanova:97}, but, as we recall, their
  estimates were based on much smaller systems.

\begin{figure}[!hbt]
    \begin{minipage}{0.49\textwidth}
      \begin{center}
	\includegraphics[width=0.99\textwidth]{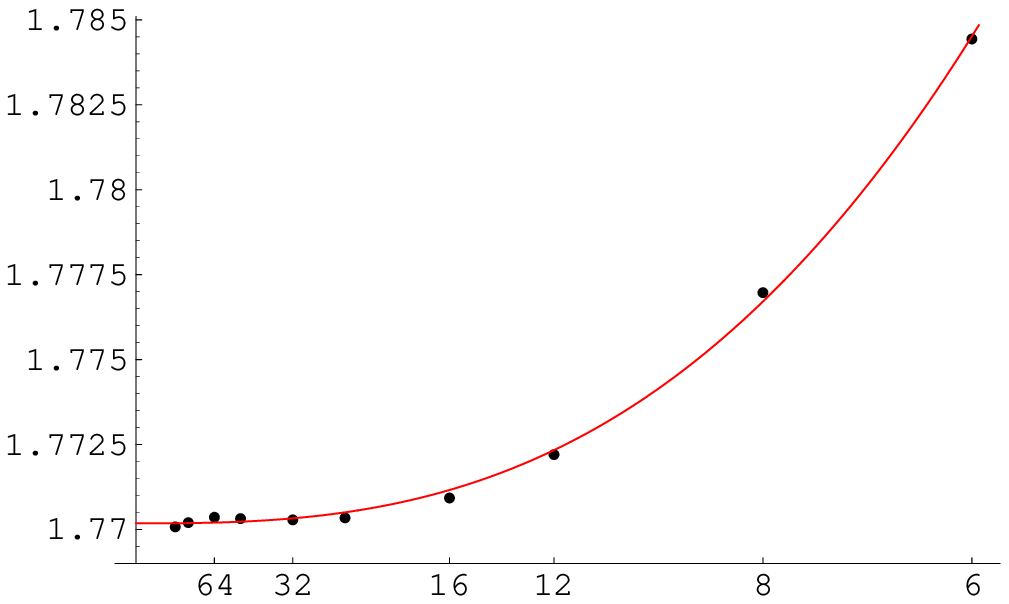}
      \end{center}
    \end{minipage}%
    \begin{minipage}{0.49\textwidth}
      \begin{center}
	\includegraphics[width=0.99\textwidth]{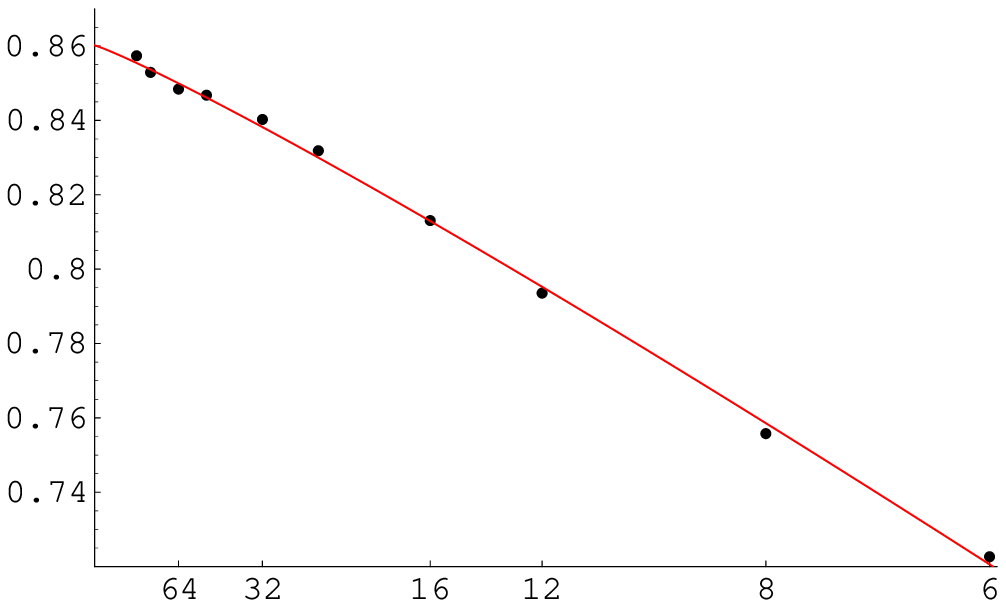}
      \end{center}
    \end{minipage}
    \caption{Free energy $\FF(K^*)$ (left) and entropy $\FS(K^*)$
      (right) vs $1/L$ together with fitted curves.}
    \label{fig:scale-2}
  \end{figure}

  Applying this procedure to the free energy, where
  $\FF(K^*)\to\FF_c$, and the entropy, where $\FS(K^*)\to\FS_c$, we
  obtained
  \begin{align*}
  \FF_c &= 1.77018 \pm 0.00005 \\
  \FS_c &= 0.86020 \pm 0.0015
  \end{align*}
  Note the considerably larger error in $\FS_c$ but also that the value
  is consistent with taking 
  \[
  \FS_c = \FF_c - 3\,K_c\,\FU_c = 0.85983 \pm 0.0025
  \]
  though we receive a slightly larger error estimate.  The data points
  and the fitted curves are shown in Figure~\ref{fig:scale-2}.
  
  For the magnetisations $\mu^-$, $\mu^+$ and $\mu^*$ we should see a
  behaviour analogous to that of the energies and we have treated them
  as such. We assume that $\mu^-\to\mu^-_c=1/3$, $\mu^*\to\mu_c$ and
  $\mu^+\to\mu^+_c$. Continuing with our trusted approach we received
  \begin{gather*}
    \mu^+_c = 0.3986 \pm 0.0015 \\
    \mu_c = 0.3711 \pm 0.0027
  \end{gather*}
  and the fitted curves are shown with the data points in 
  Figure~\ref{fig:scale-3}.
  \label{fig:scale-3}

\bibliographystyle{amsalpha} 
\bibliography{papers}

\end{document}